\documentclass[journal,12pt,onecolumn,draftclsnofoot]{IEEEtran}

\usepackage{amsfonts}
\usepackage{slashbox}
\usepackage{amsmath}
\usepackage{multirow}
\usepackage{graphicx}
\usepackage{amsfonts}
\usepackage{amsmath,epsfig}
\usepackage{mathbbold}
\usepackage{stfloats}
\usepackage{amssymb}
\usepackage{url}
\usepackage{bm}
\usepackage{cite}
\usepackage{amsmath}
\usepackage{amssymb}
\usepackage{latexsym}
\usepackage{multirow}
\usepackage{epsfig}
\usepackage{graphics}
\usepackage{mathrsfs}
\usepackage{algorithmic}
\usepackage{algorithm}
\usepackage{subfigure}
\usepackage{slashbox}
\usepackage[all]{xy}
\usepackage{setspace}

\newenvironment{shrinkeq}[1]
 { \bgroup
   \addtolength\abovedisplayshortskip{#1}
   \addtolength\abovedisplayskip{#1}
   \addtolength\belowdisplayshortskip{#1}
   \addtolength\belowdisplayskip{#1}}
 {\egroup\ignorespacesafterend}

\allowdisplaybreaks \allowdisplaybreaks[4]
\begin{document}

\title{\huge{UAV-Enabled Cooperative Jamming for Improving Secrecy of Ground Wiretap Channel} }
\author{\normalsize {An~Li,~\IEEEmembership{\normalsize{Member,~IEEE,}}
        Qingqing~Wu,~\IEEEmembership{\normalsize{Member,~IEEE,}}
        and~Rui~Zhang,~\IEEEmembership{\normalsize{Fellow,~IEEE}}}
\thanks{A. Li is with the School of Information Engineering, Nanchang University,
Nanchang, China 330031 (e-mail: lian@ncu.edu.cn).}
\thanks{Q. Wu and R. Zhang are with the Department of Electrical and Computer
Engineering, National University of Singapore, Singapore 117583 (e-mail:
\{elewuqq, elezhang\}@nus.edu.sg).}}
\maketitle
\begin{abstract}
This letter proposes a novel UAV-enabled mobile jamming scheme to improve the secrecy rate of ground wiretap channel. Specifically, a UAV is employed to transmit jamming signals to combat against eavesdropping. Such a mobile jamming scheme is particularly appealing since the UAV-enabled jammer can fly close to the eavesdropper and opportunistically jam it by leveraging the UAV's mobility. We aim to maximize the average secrecy rate by jointly optimizing the UAV's trajectory and jamming power over a given flight period. To make the problem more tractable, we drive a closed-form lower bound for the achievable secrecy rate, based on which the UAV's trajectory and transmit power are optimized alternately by an efficient iterative algorithm applying the block coordinate descent and successive convex optimization techniques. Simulation results demonstrate that the proposed joint design can significantly enhance the secrecy rate of the considered wiretap system as compared to benchmark schemes.
\end{abstract}
\vspace{-0.5cm}
\vspace{-0.45cm}
\section{Introduction}
Guarantying the secrecy of wireless communications is a critical issue due to the broadcast and shared nature of wireless channels. Cooperation based physical layer security has emerged as a promising solution to improve the secrecy of single-antenna communication systems\cite{Bassily2013}. One of the most common cooperative techniques for physical layer security is cooperative jamming (see \cite{Cumanan2017,Lai2008} and the references therein), where friendly jammers are employed to collaboratively transmit interfering signals to weaken the quality of the wiretap channel and hence enhance the secrecy rate. However, conventional static jamming schemes assumed that the locations of ground jammers are fixed or quasi-static, thus giving rise to the following two major challenges. First, the static jammers are not helpful when they are far away from the eavesdroppers, and even decrease the secrecy rate when they are close to the destination. Second, the perfect instantaneous channel state information (CSI) of jammer-eavesdropper link is generally required to perform effective jamming. However, the randomness of terrestrial wireless channels (e.g., shadowing and small-scale fading) not only degrades the jamming performance, but also makes it difficult and even impossible to obtain accurate CSI in practice, especially when the eavesdropper is passive.

Recently, unmanned aerial vehicles (UAVs) have been increasingly applied in wireless communications\cite{Zeng2016}, such as UAV-mounted BSs \cite{Wu2017,JR:wu2017_ofdm,JR:wu2017_capacity}, UAV-enabled relaying\cite{Zenga2016}, and UAV-aided data collection/dissemination  due to their many advantages such as cost-effective deployment, controllable mobility, and line-of-sight (LoS) air-to-ground link. 
All these features provide new opportunities to use UAVs as mobile jammers to tackle the above two critical issues in conventional cooperative jamming for ground wiretap channels. First, subject to practical mobility constraints on the initial/final locations as well as the maximum speed, a UAV employed as a mobile jammer can opportunistically interfere with potential eavesdroppers on the ground with more jamming power when it comes closer to each of the eavesdroppers and is sufficiently distant away from the destination, which helps enhance the jamming performance. Second, the LoS channel from the UAV to each ground eavesdropper brings the following two benefits as compared to terrestrial wireless channels. One is that the channel power gain between a UAV and an eavesdropper can be easily obtained since it only depends on their distance. Note that the eavesdropper's location can be practically detected via a UAV-mounted camera or radar. Furthermore, the channel is significantly less impaired by terrestrial fading and shadowing, thus making the jamming more effective.

Motivated by the above benefits, we consider in this letter a UAV-enabled mobile jammer for improving the secrecy rate of a ground three-terminal wiretap channel. Specifically, subject to both average and peak transmit power constraints as well as the UAV's mobility constraints, a joint UAV trajectory design and power control scheme is proposed to maximize a derived lower bound of the achievable secrecy rate over a finite UAV flight period. To tackle the non-convexity of the considered optimization problem, an efficient iterative algorithm is proposed by applying the block coordinate descent and successive convex optimization techniques to find a high-quality approximate solution. Numerical results verify that the proposed joint design achieves significant secrecy rate gain as compared to benchmark schemes without power control or trajectory optimization. Notice that a secrecy UAV communication system has been recently studied in\cite{Zhang2017}, while its difference from this letter lies in that the UAV is considered as the legitimate source in\cite{Zhang2017} instead of a cooperative jammer as in this letter.

\vspace{-0.35cm}
\section{System Model}
As shown in Fig.1, we consider a three-terminal ground wiretap system where a source S transmits information to a destination D in the presence of an eavesdropper E. All ground nodes are assumed at fixed locations which are known {\it a priori}. To improve the secrecy rate from S to D, a UAV is employed as a mobile jammer to cooperatively transmit jamming signals to combat against the eavesdropping by E over a given flight period $T$ in second (s). Intuitively, a larger period $T$ in general provides the UAV more time to move closer to E to impose stronger jamming while keeping farther away from D to cause less interference, and hence helps achieve a higher secrecy rate. 
\begin{figure}[t]
 \centering
   \includegraphics[scale=0.35]{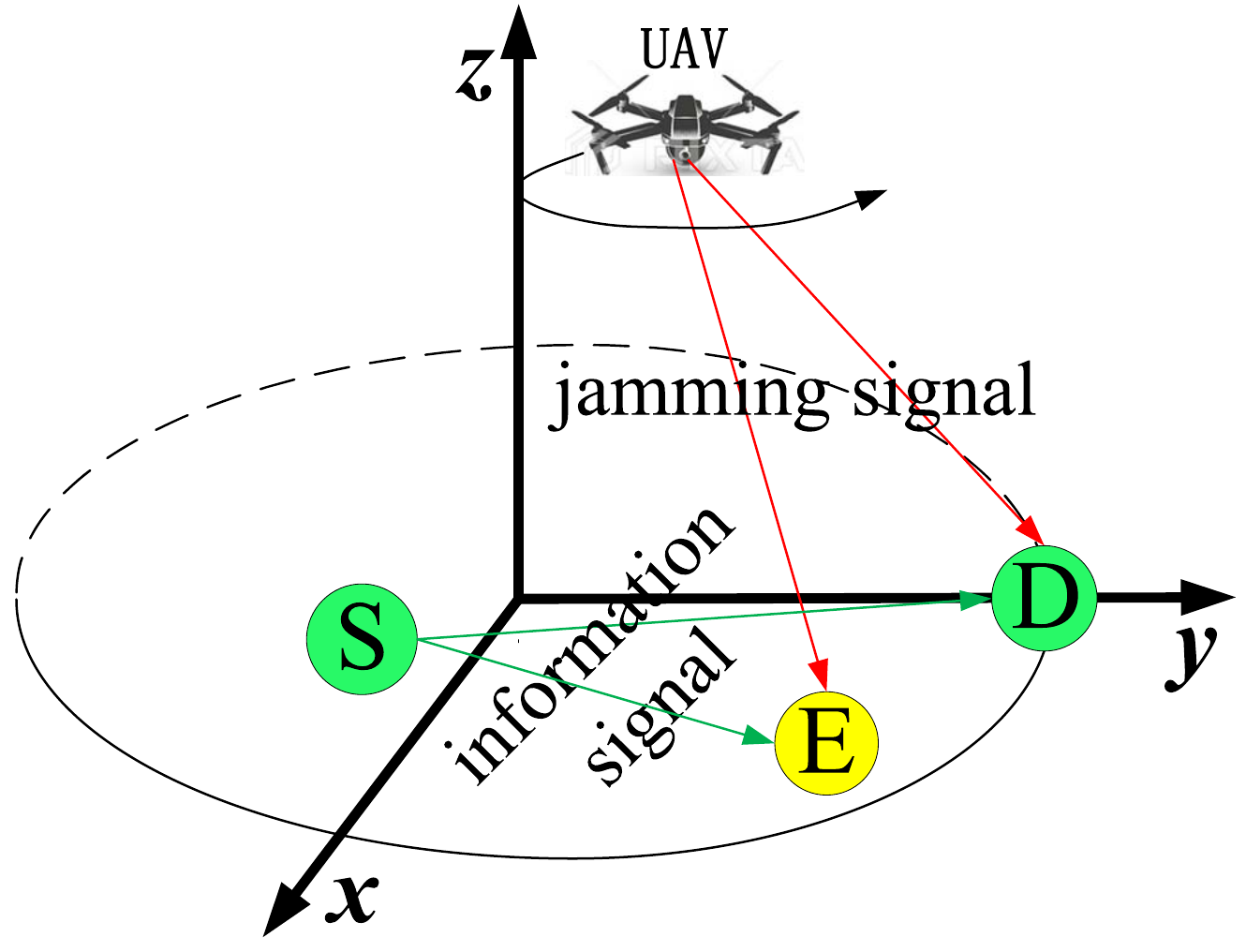}
   \vspace{-0.3cm}
   \caption{\small{ A UAV-enabled cooperative jamming system.}}
 \label{Figure 1}
 \vspace{-0.75cm}
\end{figure}

Without loss of generality, we consider a three-dimensional (3D) Cartesian coordinate system with the ground user $i$'s horizontal coordinate denoted by $\mathbf{w}_i=[x_i,y_i]^T$ in meter (m), $i\in$ \{S, D, E\}. It is assumed that the UAV flies horizontally at a constant altitude $H$ in m and its initial/final horizontal locations, denoted by $\mathbf q_0$ and $\mathbf q_F$ respectively, are pre-determined depending on its take-off/landing sites or specific mission requirement. Similar to\cite{Zenga2016}, the UAV's flight period $T$ is discretized into $N$ equal-length time slots each with duration $\delta_t = T/N$ whereby the UAV's trajectory over $T$ can be approximated by a length-$N$ sequence $\mathbf q[n]=[x[n], y[n]]^T, n\in\mathcal{N}=\{1,\cdots,N\}$, which satisfies the following mobility constraints:
\begin{shrinkeq}{-1.2ex}
\begin{subequations}\label{eq:1}
\begin{align}
& \|\mathbf q[n+1]-\mathbf q[n]\|^2 \leq L^2, n = 1, \cdots, N-1, \\
& \|\mathbf q[1]-\mathbf q_0\|^2 \leq L^2, \mathbf q[N]=\mathbf q_F,
\vspace{-0.25cm}
\end{align}
\end{subequations}
\end{shrinkeq}
where $L = V\delta_t$ is the maximum horizontal distance that the UAV can fly within each time slot assuming its maximum speed is $V$ in m/s. Notice that $N$ (or $\delta_t$) needs to be chosen sufficiently large (small) such that $L$ is small enough compared with $H$ to ensure that the UAV-ground channels are approximately constant within each slot.

We assume that the UAV-ground channels are mainly dominated by the LoS link\cite{Wu2017 ,Zenga2016}. Thus, the channel power gain at time slot $n$ follows the free-space path loss model as
\vspace{-0.15cm}
\begin{equation}\label{eq:2}
h_{i}[n] = \rho_0 d_{\text{U}i}^{-2}[n]=\frac{\rho_0}{\|\mathbf q[n]-\mathbf{w}_i\|^2+H^2}, n\in\mathcal{N},\vspace{-0.15cm}
\end{equation}
where $d_{\text{U}i}[n], i \in$ \{\text{D, E}\} is the distance between the UAV and ground user $i$ in time slot $n$, and $\rho_0$ denotes the channel power gain at the reference distance $d_0 = 1$ m.

Both ground channels for the S$\rightarrow i$ links are assumed to be independent Rayleigh fading with the channel power gains denoted by $g_i = \rho_0 d_{\text{S}i}^{-\varphi}\xi_i$, $i\in \{\text{D, E}\}$, where $\varphi$ is the path loss exponent and $\xi_i$ is an independent exponentially distributed random variable with unit mean. Note that $\delta_t$ is generally much larger than the coherence time of ground channels, which are thus assumed stationary and ergodic within each slot. Let $P_\text{S}[n]$ and $P_\text{U}[n]$ denote respectively the information signal transmit power at source S and the jamming signal power by the UAV in time slot $n$. In practice, they are subject to both average and peak power constraints as follows
\begin{shrinkeq}{-1.2ex}
\begin{subequations}\label{eq:3}
\begin{align}
& \frac{1}{N}\sum_{n=1}^{N}P_\text{S}[n] \leq \bar{P}_\text{S}, \quad 0\leq P_\text{S}[n] \leq P_\text{Smax}, n\in\mathcal{N},  \\
& \frac{1}{N}\sum_{n=1}^{N}P_\text{U}[n] \leq \bar{P}_\text{U},\quad 0\leq P_\text{U}[n] \leq P_\text{Umax}, n\in\mathcal{N},
\vspace{-0.15cm}
\end{align}
\end{subequations}
\end{shrinkeq}
where $\bar{P}_\text{S} \leq P_\text{Smax}$ and $\bar{P}_\text{U} \leq P_\text{Umax}$.
Thus, the average achievable secrecy rate in bits/second/Hertz (bps/Hz) over $N$ time slots is given by\cite{Gopala2008}
\begin{shrinkeq}{-1.2ex}
\begin{equation}\label{eq:4}
R = \frac{1}{N}\sum_{n=1}^{N}\left[R_\text{D}[n] - R_\text{E}[n]\right]^+,
\end{equation}
\end{shrinkeq}
with $[x]^+ \triangleq \text{max}(x,0)$. $R_\text{D}[n] = \mathbb E[\text{log}_2(1+\frac{g_\text{D}P_\text{S}[n]}{h_\text{D}[n]P_\text{U}[n]+\sigma^2})]$, $R_\text{E}[n]=\mathbb E[\text{log}_2(1+ \frac{g_\text{E}P_\text{S}[n]}{h_\text{E}[n]P_\text{U}[n]+\sigma^2})]$, where $\mathbb E[\cdot]$ is the expectation operator with respect to ground fading channels, and $\sigma^2$ is the independent Gaussian noise power at D or E.

\vspace{-0.3cm}
\section{Problem Formulation}
Let $\mathbf Q = \{\mathbf q[n], n\in\mathcal{N}\}$, $\mathbf P_\text{S}=\{P_\text{S}[n],n\in\mathcal{N}\}$, and $\mathbf P_\text{U}=\{P_\text{U}[n],n\in\mathcal{N}\}$. Our objective is to maximize the average achievable secrecy rate $R$ in (\ref{eq:4}) by jointly optimizing the UAV's trajectory $\mathbf Q$ and the transmit power $\mathbf P_\text{S}$ and $\mathbf P_\text{U}$ over all time slots subject to UAV's mobility constraints in (\ref{eq:1}) and transmit power constraints in (\ref{eq:3}). Thus, the optimization problem can be formulated as
\vspace{-0.25cm}
\begin{align}
(\text P1): \max \limits_{\mathbf{Q, P}_\text{S}, \mathbf P_\text{U}} & \sum_{n=1}^{N}(R_\text{D}[n] - R_\text{E}[n])   \\
\text{s.t.} \qquad &(\ref{eq:1}), (\ref{eq:3}), \nonumber
\end{align}
where the operation $[\cdot]^+$ is omitted since each summation term in the objective function of (P1) must be non-negative at the optimal solution; otherwise, the optimal value of (P1) can be increased by setting $P_\text{S}[n]=0$ for any such $n$ without violating the power constraints. Note that (P1) is still difficult to solve due to its non-convex objective function with respect to $\mathbf{Q}$, $\mathbf{P}_\text{S}$, and $\mathbf P_\text{U}$. To simplify the problem, we derive a lower bound for the objective value (achievable secrecy rate) of (P1), where $R_\text{D}[n]$ and $R_\text{E}[n]$ are replaced by their lower and upper bounds, respectively.

Based on the convexity of $\text{ln}(1+e^{x})$ and Jensen's inequality, $R_\text{D}[n]$ is lower-bounded by
\begin{shrinkeq}{-1.2ex}
\begin{equation}\label{eq:7}
\begin{split}
R_\text{D}[n] & = \frac{1}{\text{ln}2}\mathbb E\left[ \text{ln}\left(1+X_n \right)\right] = \frac{1}{\text{ln}2}\mathbb E\left[ \text{ln}\left(1+e^{\text{ln}X_n} \right)\right] \\
& \geq \frac{1}{\text{ln}2} \text{ln}\left( 1 + e^{\mathbb E[\text{ln}X_n]}\right),
\end{split}
\end{equation}
\end{shrinkeq}
where $X_n=a_ng_\text{D}$ with $a_n=\frac{P_\text{S}[n]}{\frac{\rho_0P_\text{U}[n]}{\|\mathbf q[n]-\mathbf{w}_\text{D}\|^2+H^2}+\sigma^2}$.
%
%
Since $X_n$ is an exponential distributed random variable with parameter $\lambda_n = \frac{1}{\rho_0a_n}d_\text{SD}^{\varphi}$, we obtain by using eq.(4.331.1) in \cite{IS2007}
\vspace{-0.3cm}
\begin{equation}\label{eq:8}
\mathbb E[\text{ln}X_n] = \int_0^{\infty}\text{ln}x\lambda_n e^{-\lambda_n x}dx = -\text{ln}\lambda_n-\kappa, \vspace{-0.15cm}
\end{equation}
where $\kappa$ is the Euler constant. Substituting (\ref{eq:8}) into (\ref{eq:7}), the lower bound $R_\text{D}^\text{lo}[n]$ of $R_\text{D}[n]$ is given by
\vspace{-0.2cm}
\begin{equation}\label{eq:9}
R_\text{D}[n] \geq R_\text{D}^\text{lo}[n] = \text{log}_2\left(1+\frac{e^{-\kappa}\gamma_0d_\text{SD}^{-\varphi}P_\text{S}[n]}{\frac{\gamma_0P_\text{U}[n]}{\|\mathbf q[n]-\mathbf{w}_\text{D}\|^2+H^2}+1} \right), \vspace{-0.15cm}
\end{equation}where $\gamma_0=\frac{\rho_0}{\sigma^2}$. Due to the concavity of the function $\text{ln}(1+x)$, an upper bound $R_\text{E}^\text{up}[n]$ of $R_\text{E}[n]$ is given by
\vspace{-0.3cm}
\begin{equation}\label{eq:10}
R_\text{E}[n]\leq R_\text{E}^\text{up}[n] = \text{log}_2\left(1+\frac{\gamma_0d_\text{SE}^{-\varphi}P_\text{S}[n]}{\frac{\gamma_0P_\text{U}[n]}{\|\mathbf q[n]-\mathbf{w}_\text{E}\|^2+H^2}+1}\right). \vspace{-0.3cm}
\end{equation}

With (\ref{eq:9}) and (\ref{eq:10}), (P1) can be approximately transformed to the following problem,
\vspace{-0.25cm}
\begin{align}\label{eq:11}
(\text P2): \max \limits_{\mathbf{Q, P}_\text{S}, \mathbf P_\text{U}} & \sum_{n=1}^{N}(R_\text{D}^\text{lo}[n] - R_\text{E}^\text{up}[n])   \\
\text{s.t.} \quad &(1), (3).  \nonumber
\end{align}Although more tractable, problem (P2) is still non-convex with respect to $\mathbf{Q}$, $\mathbf{P}_\text{S}$, and $\mathbf P_\text{U}$ and difficult to be optimally solved. Thus, we propose an efficient iterative algorithm to obtain a suboptimal solution for it in the next section.
\vspace{-0.3cm}
\section{Proposed Algorithm}
 In this section, we apply block coordinate descent and successive convex optimization to (P2), which leads to an efficient iterative algorithm. Specifically, problem (P2) is partitioned into three subproblems to optimize the transmit power $\mathbf P_\text{S}$ and $\mathbf P_\text{U}$ as well as the UAV trajectory $\mathbf Q$ alternately in an iterative manner until the algorithm converges.

\vspace{-0.3cm}
\subsection{Subproblem 1: Transmit Power $\mathbf P_\text{S}$ Optimization}
For any given UAV trajectory $\mathbf Q$ and transmit power $\mathbf P_\text{U}$, problem (P2) can be written as
\vspace{-0.25cm}
\begin{align}
(\text P3): \max \limits_{\mathbf{P}_\text{S}} & \sum_{n=1}^{N}\left[\text{log}_2\left(1 + a_nP_\text{S}[n]\right) - \text{log}_2\left(1+ b_nP_\text{S}[n]\right)\right]  \\
\text{s.t.} \quad & (3\text{a}) \nonumber,
\end{align}where $a_n = \frac{e^{-\kappa}\gamma_0d_\text{SD}^{-\varphi}}{\frac{\gamma_0P_\text{U}[n]}{\|\mathbf q[n]-\mathbf{w}_\text{D}\|^2+H^2}+1}$, $b_n = \frac{\gamma_0d_\text{SE}^{-\varphi}}{\frac{\gamma_0P_\text{U}[n]}{\|\mathbf q[n]-\mathbf{w}_\text{E}\|^2+H^2}+1}$.
Although (P3) is non-convex, its optimal solution can be expressed as \cite{Gopala2008}
\vspace{-0.2cm}
\begin{equation}\label{eq:13}
P_\text{S}^*[n] =
\begin{cases}
\text{min}([\hat{P}_\text{S}[n]]^+, P_\text{Smax}) & a_n > b_n, \\
0 & a_n \leq b_n,
\end{cases}
\end{equation}
where \vspace{-0.2cm}
\begin{equation}\label{eq:14}
\hat{P}_\text{S}[n]\!=\!\sqrt{\left(\frac{1}{2b_n} - \frac{1}{2a_n}\right)^2\!+\!\frac{1}{\mu\text{ln}2}\left(\frac{1}{b_n} - \frac{1}{a_n} \right)}-\frac{1}{2a_n}-\frac{1}{2b_n},
\end{equation}where $\mu$ is a non-negative parameter ensuring $\sum_{n=1}^{N}P_\text{S}^*[n] \leq N\bar{P}_\text{S}$, which can be found efficiently via the bisection method.

\vspace{-0.3cm}
\subsection{Subproblem 2: Transmit Power $\mathbf P_\text{U}$ Optimization}
Let $c_n=e^{-\kappa}\gamma_0d_\text{SD}^{-\varphi}P_\text{S}[n]$, $d_n=\frac{\gamma_0}{\|\mathbf q[n]-\mathbf{w}_\text{D}\|^2+H^2}$, $e_n=\gamma_0d_\text{SE}^{-\varphi}P_\text{S}[n]$, and $f_n=\frac{\gamma_0}{\|\mathbf q[n]-\mathbf{w}_\text{E}\|^2+H^2}$. For any given UAV trajectory $\mathbf Q$ and transmit power $\mathbf P_\text{S}$, problem (P2) is reformulated as
\vspace{-0.2cm}
\begin{align}\label{eq:15}
(\text P4): \max \limits_{\mathbf{P}_\text{U}} & \sum_{n=1}^{N}\left[\text{log}_2\left(1\!+\!\frac{c_n}{d_nP_\text{U}[n]+1}\right)\!-\!\text{log}_2\left(1\!+\!\frac{e_n}{f_nP_\text{U}[n]+1}\right)\right]  \\
\text{s.t.} \quad & (3\text{b}). \nonumber
\end{align}Although the objective function of (P4) is non-convex, it is the difference of two convex functions with respect to $P_\text{U}[n]$. This thus motivates us to apply the successive convex optimization technique to tackle the non-convexity of (P4) and obtain an approximate solution. Define $\mathbf{P}_\text{U}^k=\{P_\text{U}^k[n], n\in\mathcal{N}\}$ as the given UAV transmit power in the $k$-th iteration. Since the first term in (\ref{eq:15}) is a convex function of $P_\text{U}[n]$, its first-order Taylor expansion at $P_\text{U}^k[n]$ is a global under-estimator\cite{Zenga2016,Wu2017}, i.e.,
\vspace{-0.2cm}
\begin{equation}\label{eq:16}
\text{log}_2\left(1+\frac{c_n}{d_nP_\text{U}[n]+1}\right) \geq A^k[n](P_\text{U}[n]-P_\text{U}^k[n])+B^k[n], \vspace{-0.2cm}
\end{equation}where $A^k[n]=\frac{-c_nd_n}{\text{ln}2(d_nP_\text{U}^k[n]+1)(d_nP_\text{U}^k[n]+c_n+1)}$ and $B^k[n]=\text{log}_2\left(1+\frac{c_n}{d_nP_\text{U}^k[n]+1}\right)$.
With (\ref{eq:16}), problem (P4) is approximated as the following problem for any given local point $\mathbf{P}_\text{U}^k$,
\vspace{-0.2cm}
\begin{align}\label{eq:17}
(\text P5): \max \limits_{\mathbf{P}_\text{U}} & \sum_{n=1}^{N}\left[A^k[n]P_\text{U}[n]-\text{log}_2\left(1+\frac{e_n}{f_nP_\text{U}[n]+1}\right)\right]  \\
\text{s.t.} \quad & (3\text{b}). \nonumber
\end{align}Note that (P5) is a convex optimization problem and can be solved efficiently by standard convex optimization solvers such as CVX~\cite{Grant2016}. Since the first-order Taylor expansion in (\ref{eq:16}) suggests that the objective value of (P4) at $\mathbf{P}_\text{U}^k$ is the same as that of (P5), and (P5) maximizes the lower bound of the objective function of its original problem (P4), the objective value of (P4) with the solution obtained by solving (P5) is always no less than that with any $\mathbf{P}_\text{U}^k$.

\vspace{-0.35cm}
\subsection{Subproblem 3: UAV Trajectory $\mathbf Q$ Optimization}
For any given transmit power $\mathbf P_\text{S}$ and $\mathbf P_\text{U}$, by introducing slack variables $\mathbf L =\{l[n]=\|\mathbf q[n]-\mathbf{w}_\text{D}\|^2+H^2,n\in\mathcal{N}\}$ and $\mathbf M =\{m[n]=\|\mathbf q[n]-\mathbf{w}_\text{E}\|^2+H^2,n\in\mathcal{N}\}$, (P2) can be written as
\vspace{-0.25cm}
\begin{subequations}\label{eq:18}
\begin{align}
(\text P6): \max \limits_{\mathbf{Q,L,M}} & \sum_{n=1}^{N}\left[\text{log}_2\left(1\!+\!\frac{c_n}{\frac{\gamma_0P_\text{U}[n]}{l[n]}+1}\right)\!-\!\text{log}_2\left(1\!+\!\frac{e_n}{\frac{\gamma_0P_\text{U}[n]}{m[n]}+1}\right)\right]  \\
\text{s.t.} \quad & l[n]-\|\mathbf q[n]-\mathbf{w}_\text{D}\|^2-H^2 \leq 0, \\
& \|\mathbf q[n]-\mathbf{w}_\text{E}\|^2+H^2-m[n] \leq 0, \\
& (1). \nonumber
\end{align}
\end{subequations}It can be verified that at the optimal solution to problem (P6), constraints (\ref{eq:18}b) and (\ref{eq:18}c) must hold with equalities, since otherwise $l[n]$ ($m[n]$) can be increased (decreased) to improve the objective value. Similarly, to handle the non-convexity of (\ref{eq:18}a) and (\ref{eq:18}b) with respect to $m[n]$ and $\mathbf q[n]$, respectively, the successive convex optimization technique is applied where the terms $\text{log}_2\left(1+\frac{e_nm[n]}{m[n]+\gamma_0P_\text{U}[n]}\right)$ and $-\|\mathbf q[n]-\mathbf{w}_\text{D}\|^2$ are replaced by their respective concave upper bound at a given local point. Define $\mathbf Q^k=\{\mathbf q^k[n],n\in\mathcal{N}\}$ as a given initial trajectory in the $k$-th iteration; then we obtain
\vspace{-0.25cm}
\begin{subequations}\label{eq:19}
\begin{align}
\text{log}_2\left(1+\frac{e_n}{\frac{\gamma_0P_\text{U}[n]}{m[n]}+1}\right) & \leq C^k[n](m[n]-m^k[n])+F^k[n], \\
-\|\mathbf q[n]-\mathbf{w}_\text{D}\|^2 & \leq G^k[n],
\end{align}
\end{subequations}
where \quad $C^k[n]=\frac{e_n\gamma_0P_\text{U}[n]}{\text{ln}2(m^k[n]+\gamma_0P_\text{U}[n])((e_n+1)m^k[n]+\gamma_0P_\text{U}[n])}$, \quad $m^k[n]=\|\mathbf q^k[n]-\mathbf{w}_\text{E}\|^2$, \quad $F^k[n]=\text{log}_2\left(1+\frac{e_nm^k[n]}{m^k[n]+\gamma_0P_\text{U}[n]}\right)$, and $G^k[n]=\|\mathbf q^k[n]\|^2-2[\mathbf q^k[n]-\mathbf{w}_\text{D}]^T\mathbf q[n]-\|\mathbf{w}_\text{D}\|^2$.

With (\ref{eq:19}), problem (P6) is recast as
\vspace{-0.25cm}
\begin{subequations}\label{eq:20}
\begin{align}
(\text P7): \max \limits_{\mathbf{Q,L,M}} & \sum_{n=1}^{N}\left[\text{log}_2\left(1+\frac{c_n}{\frac{\gamma_0P_\text{U}[n]}{l[n]}+1}\right)-C^k[n]m[n]\right]  \\
\text{s.t.} \quad & l[n]+G^k[n]-H^2 \leq 0, \\
& (\ref{eq:18}\text{c}),(1). \nonumber
\end{align}
\end{subequations}Since (P7) is a convex optimization problem, it can be efficiently solved by CVX. Similarly, the upper bounds adopted in (\ref{eq:19}) guarantee the feasible set of (P7) to be a feasible subset of (P6). As such, the objective value of (P6) with the solution obtained from (P7) is always no less than that with any $\mathbf Q^k$.

\vspace{-0.5cm}
\subsection{Overall Algorithm}
In summary, the proposed algorithm solves three subproblems (P3), (P5), and (P7) alternately in an iterative manner by applying the block coordinate descent method until the fractional increase of the objective value is below a given small threshold, $\epsilon>0$. As illustrated in Subsections A-C, the objective values of (P2) with the solutions by solving the subproblems (P3), (P5), and (P7) are non-decreasing over iterations. Since the objective value of (P2) is finite, the proposed iterative algorithm is guaranteed to converge.
\vspace{-0.4cm}
\section{Numerical Results}
To demonstrate the performance of the proposed joint trajectory and power control design (denoted by ``J-T\&P"), we compare it with two benchmark algorithms: trajectory optimization without power control ``T/NP" and line-segment trajectory with optimized power control ``LT/P". Specifically, in ``T/NP", the transmit power of the UAV or S in each slot is set as their corresponding average power, and the UAV's trajectory is optimized by solving problem (P7) iteratively until convergence. In ``LT/P", the UAV's trajectory is designed in a best-effort manner: the UAV firstly flies towards the location above E, then hovers above E, and finally flies at the maximum speed to reach its final location by the last time slot. Note that if $T$ is not sufficiently large for the UAV to reach E, the UAV will turn at a certain midway point then fly towards its final location at the maximum speed. Therefore, for ``LT/P", the pre-determined trajectory consists of two line segments, and the power control is obtained by alternately solving subproblems 1 and 2. The parameters are set as follows: $\mathbf{q}_0=[-100,100]^T$ m, $\mathbf{q}_F=[500,100]^T$ m, $H=100$ m, $V = 3$ m/s, $\mathbf{w}_\text{S}=[0,0]^T$ m, $\mathbf{w}_\text{D}=[300,0]^T$ m, $\mathbf{w}_\text{E}=[200,200]^T$ m, $\gamma_0=90$ dB, $\bar{P}_\text{U}=10$ dBm, $P_\text{Umax}=4\bar{P}_\text{U}=16$ dBm, $\bar{P}_\text{S}=30$ dBm, $P_\text{Smax}=36$ dBm, and $\epsilon=10^{-4}$.

Fig.2(a) shows the UAV's trajectories versus the period $T$. The source S, destination D, eavesdropper E, and the UAV's initial and final locations are marked with $\bigcirc, \square, \times, +$, and $\ast$, respectively. It is observed that when $T=200$ s, which is the minimum required time for the UAV to fly from the initial location to the final location at the maximum speed, the trajectories of the ``J-T\&P", ``LT/P", and ``T/NP" algorithms are identical. However, their trajectories appear gradually different as $T$ increases. In particular, when $T=350$ s, significant trajectory differences can be observed for the three algorithms. Specifically, for ``T/NP", it is observed that the UAV flies along the outermost trajectory and thus spends more time on travelling than that in ``J-T\&P", whereas for ``LT/P", the UAV takes the shortest travelling time. This is because for ``T/NP", the power control is not considered and thus the UAV tends to keep as far away as possible to avoid causing excessive interference to D. However, for the proposed ``J-T\&P", the UAV is able to decrease (increase) the jamming power when it flies closer to (farther away from) D. Further, it is observed that for all algorithms, the UAV first reaches a certain location (not directly above E for ``J-T\&P" and ``T/NP"), then remains stationary at this location as long as possible, and finally reaches the final location by the last time slot. This is because these hovering locations generally strike an optimal balance between degrading the wiretap channel and causing undesired interference to the destination and hence achieve the maximum secrecy rate in each case.
\begin{figure}
\subfigure[UAV's trajectories ]{\begin{minipage}{0.5\linewidth}
{\includegraphics[width=2.6in]{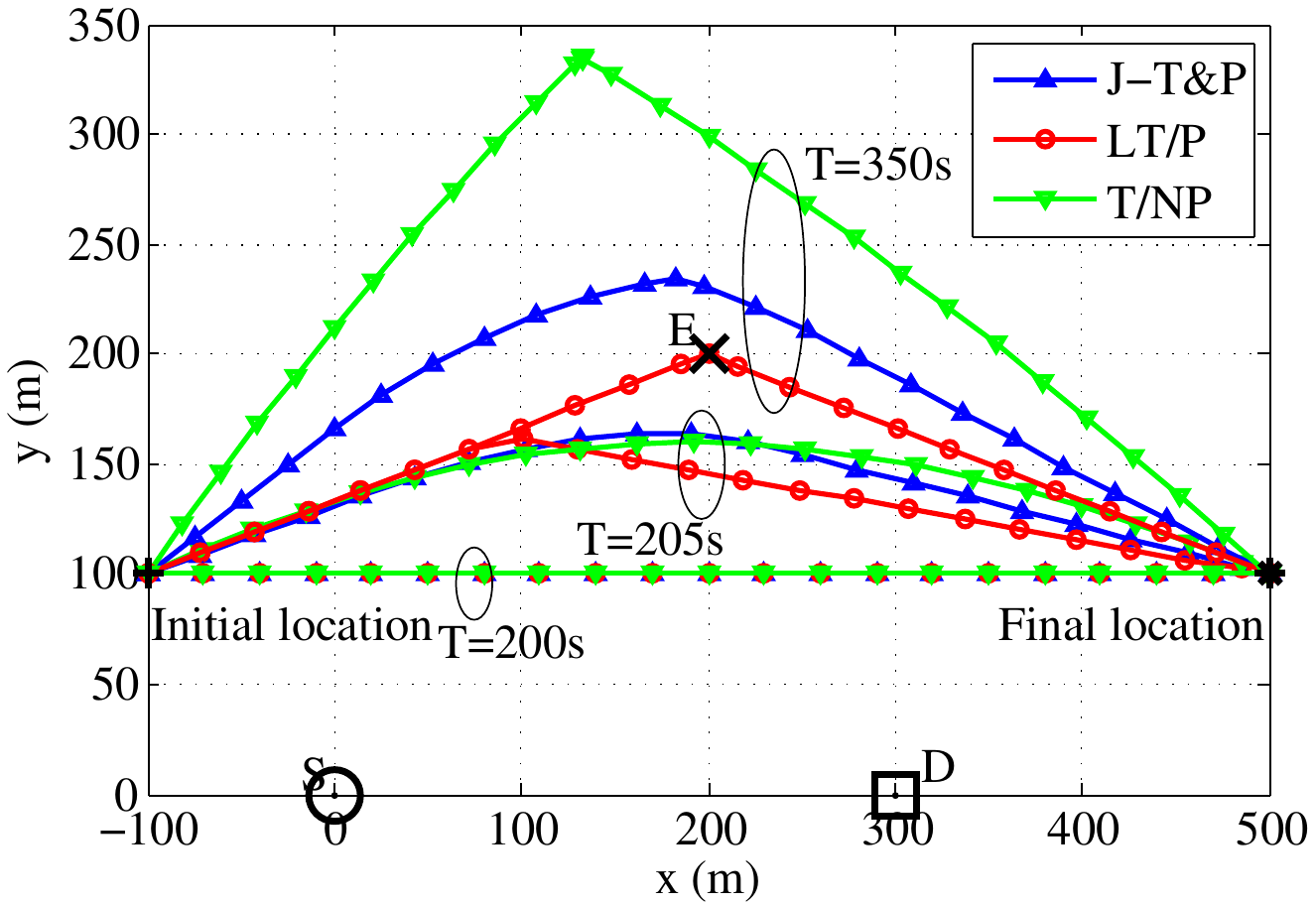}}
\end{minipage}}~~\subfigure[Secrecy rate versus $T$]{\begin{minipage}{0.5\linewidth} {\includegraphics[width=2.5
in]{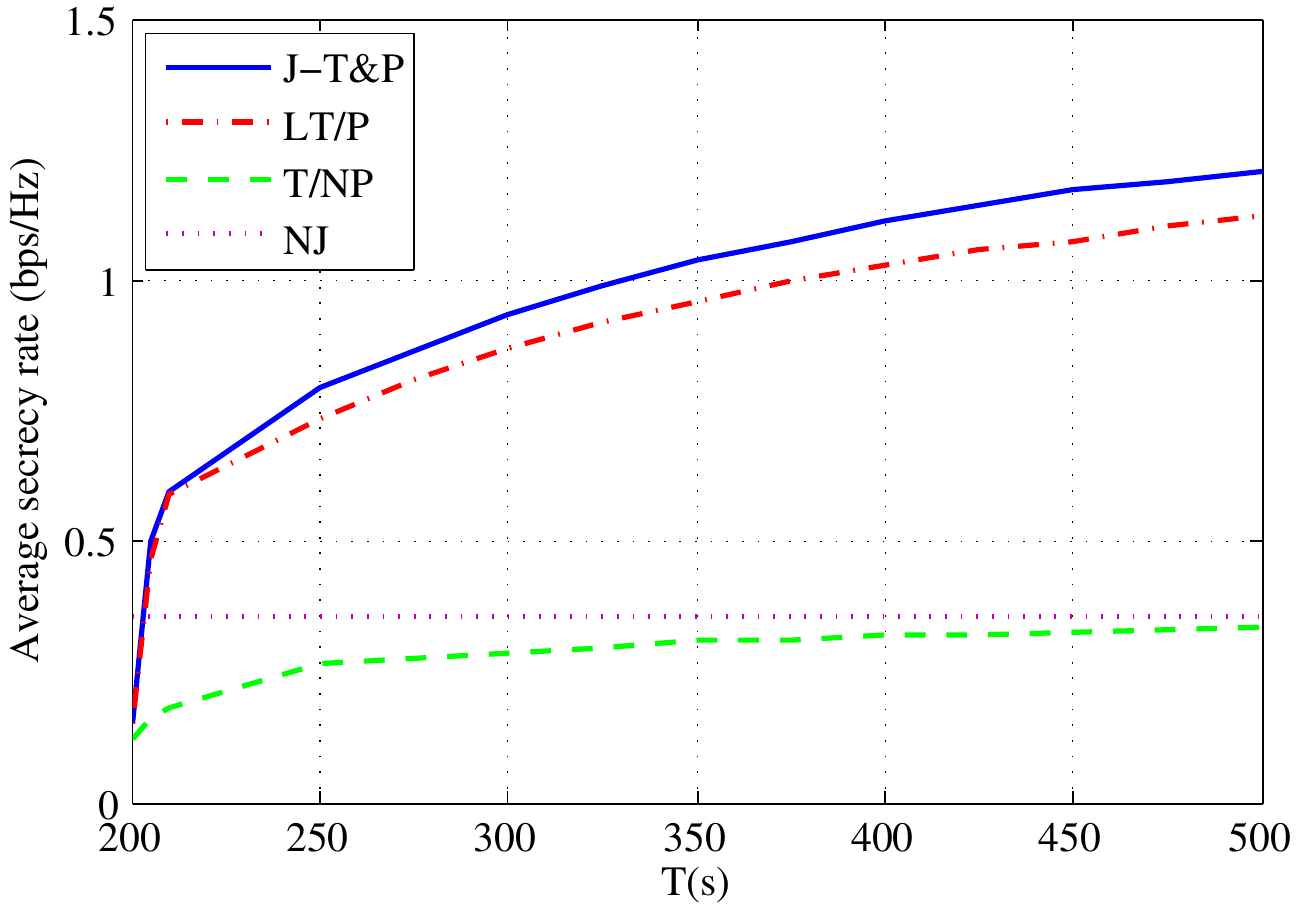}}
\end{minipage}} \vspace{-0.2in}
\caption{Trajectories of UAV-enabled jammer and achievable secrecy rates.}
\label{fig:graph2} \vspace{-0.3in}
\end{figure}

Fig.2(b) shows the average achievable secrecy rate versus $T$ where the scheme without a jammer (denoted by ``NJ") is also considered for comparison. It is observed that the secrecy rates achieved by all algorithms except ``NJ" increase as $T$ increases, as expected. Besides, it is observed that the proposed ``J-T\&P" algorithm always achieves the highest secrecy rate while the benchmark ``T/NP" achieves even lower secrecy rate than ``NJ". Such results validate the necessity of joint UAV trajectory and power control design for mobile jamming.
\vspace{-0.2cm}
\section{Conclusion}
In this letter, a mobile UAV-enabled jammer is employed to opportunistically jam the eavesdropper, thus improving the secrecy rate of the ground wiretap channel. Specifically, an efficient iterative algorithm is proposed to maximize the achievable average secrecy rate over a given finite period, subject to the average and peak transmit power constraints as well as the UAV's mobility constraints. Numerical results show that jointly optimizing the UAV's trajectory with source/UAV transmit power can significantly enhance the physical layer security performance of ground wiretap channels.


\vspace{-0.4cm}
\begin{thebibliography}{1}

\bibitem{Bassily2013}
R. Bassily, et al., ``Cooperative security at the physical layer: a summary of recent advances," {\em IEEE Signal Process. Mag.}, vol. 30, no. 5, pp. 16-28, Sep. 2013.

\bibitem{Lai2008}
L. Lai and H. E. Gamal, ``The relay-eavesdropper channel: cooperation for secrecy," {\em IEEE Trans. Inf. Theory}, vol. 54, no. 9, pp. 4005-4019, Sep. 2008.

\bibitem{Cumanan2017}
K. Cumanan, et al., ``Physical layer security jamming: theoretical limits and practical designs in wireless networks," {\em IEEE Access}, vol. 5, pp. 3603-3611, Mar. 2017.


\bibitem{Zeng2016}
Y. Zeng, R. Zhang, and T. J. Lim, ``Wireless communications with unmanned aerial vehicles: opportunities and challenges," {\em IEEE Commun. Mag.}, vol. 54, no. 5, pp. 36-42, Sep. 2008.

\bibitem{Wu2017}
Q. Wu, Y. Zeng, and R. Zhang, ``Joint trajectory and communication design for multi-UAV enabled wireless networks," {\em IEEE Trans. Wireless Commun.}, accepted, 2018.

\bibitem{JR:wu2017_ofdm}
Q.~Wu and R.~Zhang, ``Common throughput maximization in {UAV}-enabled {OFDMA}
  systems with delay consideration,'' \emph{submitted to IEEE Trans. Commun.},
  2017,  [Online] Available: https://arxiv.org/abs/1801.00444.
  
  \bibitem{JR:wu2017_capacity}
Q.~Wu, J.~Xu, and R.~Zhang, ``Capacity characterization of {UAV}-enabled
  two-user broadcast channel,'' \emph{submitted to IEEE J. Sel. Areas Commun.},
  2017, [Online] Available: https://arxiv.org/abs/1801.00443.

\bibitem{Zenga2016}
Y. Zeng, R. Zhang, and T. J. Lim, ``Throughput maximization for UAV-enabled mobile relaying systems," {\em IEEE Trans. Commun.}, vol. 64, no. 12, pp. 4983-4996, Dec. 2016.



\bibitem{Zhang2017}
G. Zhang, Q. Wu, M. Cui, and R. Zhang, ``Securing UAV communication via trajectory optimization," in {\em Proc. IEEE GLOBECOM}, Dec. 2017.
%
\bibitem{Gopala2008}
P. K. Gopala, L. Lai, and H. E. Gamal, ``On the secrecy capacity of fading channels," {\em IEEE Trans. Inf. Theory}, vol. 54, no. 10, pp. 4687-4698. Oct. 2008.

\bibitem{IS2007}
I. S. Gradshteyn and I. M. Ryzhik, {\em Table of Integrals, Series and Products, 7th edition}. San Diego, CA: Academic, 2007.

\bibitem{Grant2016}
M. Grant and S. Boyd, {\em CVX: MATLAB Software for Disciplined Convex Programming, Version 2.1}, accessed on Mar. 2016. [Online]. Available: http://cvxr.com/cvx.
\end{thebibliography}
\end{document}